\documentclass[manuscript]{acmart}
\usepackage{graphicx} 
\usepackage[inline]{enumitem}
\usepackage{textcomp}
\usepackage{longtable}
\usepackage{caption}
\usepackage[capitalize, noabbrev]{cleveref}
\usepackage[dvipsnames]{xcolor}
\usepackage{pgffor}
\usepackage{tabularx}
\usepackage{array}
\usepackage{booktabs}
\usepackage{multirow}
\newcolumntype{L}{>{\raggedright\arraybackslash}X}
\newcolumntype{P}[1]{>{\raggedright\arraybackslash}p{#1}}

\title{Interface Design to Support Legal Reading and Writing: Insights from Interviews with Legal Experts} 

\author{Chelse Swoopes}
\email{cswoopes@g.harvard.edu}
\orcid{0000-0002-7813-8937}
\affiliation{%
  \institution{Harvard University}
  \country{USA}
}

\author{Ziwei Gu}
\email{ziweigu@g.harvard.edu}
\orcid{0000-0001-9044-2651}
\affiliation{%
  \institution{Harvard University}
  \country{USA}
}

\author{Elena L. Glassman}
\email{glassman@seas.harvard.edu}
\orcid{0000-0001-5178-3496}
\affiliation{%
  \institution{Harvard University}
  \country{USA}
}

\date{August 2025}

\newif\ifshownotes
\shownotesfalse  

\ifshownotes
  \newcommand{\csnote}[1]{\textcolor{Cyan}{[CS: #1]}}
  \newcommand{\zgnote}[1]{\textcolor{Purple}{[ZG: #1]}}
  \newcommand{\egnote}[1]{\textcolor{Orange}{[EG: #1]}}
\else
  \newcommand{\csnote}[1]{}
  \newcommand{\zgnote}[1]{}
  \newcommand{\egnote}[1]{}
\fi

\begin{abstract}  
Legal professionals spend significant time reading, writing, and interpreting complex documents, yet research has not fully captured how they approach these tasks or what they expect from skimming and writing-support tools. To examine practices and views on emerging tools, we interviewed 22 legal professionals about workflows, challenges, and technology use. In each session, we leveraged prior HCI-based skimming and writing prototypes that surface emergent cross-document relationships and support AI-resilient interaction (noticing, judging, and recovering from model errors or unexpected behavior); participants completed a contextual fit evaluation to assess whether and how they would use the tools, which document types, and at what stages in their work. Our analysis details limitations and challenges in workflows, domain-specific feedback on AI-resilient interfaces, and expert insights on legal tech design. These findings offer actionable guidance for technology designers developing reading and writing-support for legal professionals, and for legal professionals seeking peer-informed tool integration strategies.

\end{abstract} 

\begin{document}

\maketitle

\section{Introduction} 
Reading, writing, and interpreting complex documents are fundamental tasks in the legal profession. Legal practitioners engage with high volumes of dense, technical texts that require close attention to detail, domain-specific expertise, and careful revision. While there is rich scholarship on legal-writing pedagogy, we know far less about day-to-day reading and writing practices, and how emerging technology, particularly AI-based tools, support or disrupt these workflows. Legal professionals face mounting pressures due to growing information volume and complexity, tighter timelines, and rising expectations for speed and accuracy~\cite{wolterskluwer2024frl,trigeorgetown2025stateuslegalmarket}. In response, they are adopting reading and writing support tools, including AI-enabled research, summarization, and drafting aids~\cite{aba2024techsurvey,ilta2024techsurveyexecsummary}. While it is common to highlight technical capabilities or generalized productivity gains, comparatively less is known about how such systems align with the situated needs, values, and constraints of legal professionals.

To address this gap, we conducted a semi-structured interview study with 22 legal professionals across a broad range of roles and experience levels. Our goal was to better understand how they approach legal reading and writing tasks, what AI-based and non-AI-based tools they currently use, and how these interfaces support or disrupt their workflows, what gaps and limitations they encounter, and how they evaluate new technologies aimed at supporting their workflows.

Rather than follow a standard need-finding procedure, e.g., a contextual inquiry or semi-structured need-finding interviews or focus groups like \citet{solovey2025interacting} conducted with members of the UK judiciary, we wished to jolt them out of their pre-existing beliefs about what AI could and could not do and present three prototypes of atypical (AI-resilient~\cite{gu2024ai}, text-preserving) AI-powered interfaces that augment reading and writing, which were only recently evaluated (in non-legal contexts): AI-supported skimming~\cite{gu2024ai}, comparative close reading at scale~\cite{abstractexplorer}, and writing with increased awareness of emergent content and style patterns in a corpus of related documents~\cite{dang2025corpusstudio}. We then invited them to tell us about their own hopes and pain points, given the capabilities they now knew existed. This aligns with longstanding human-computer interaction (HCI) methods that use speculative or exploratory tools to elicit situated user feedback, rather than assuming technological fit or value by default ~\cite{gaver1999design}.

As intended, reflecting on these specific prototypes' potential utility in legal contexts sparked more broad reflections on what legal professionals actually need, value, or reject in AI-based and non-AI-based reading and writing support tools. 
Participants also shared rich accounts of their day-to-day practices, including strategies for reading and writing legal documents, the frequency with which they currently rely on prior documents or support tools, and the various types of support they seek currently from reading and writing assistant tools. 

In summary, we contribute:
\begin{itemize}
    \item Twenty two semi-structured interviews of legal professionals about their reading and writing needs, augmented with examples of recent atypical AI-powered reading and writing assistance tools.
     \item An empirical account of current reading and writing practices, challenges, and approaches in legal work.
     \item Benefits and limitations of existing reading and writing support tools (AI- and non-AI-based) in real workflows.
    \item Expert-guided design opportunities: concrete ideas and requirements for future reading and writing support tools
\end{itemize}



\section{Background}

\subsection{AI Resiliency}
\label{sec:AI-resiliency}

Recent work has introduced the concept of \emph{AI-resilient interfaces}, which aims to help users be resilient to the omissions, hallucinations, misrepresentations, and even subjectively disliked alternatives that AI systems can choose on a user's behalf~\cite{glassman2024ai}. Unlike traditional human-AI interaction guidelines that emphasize efficient dismissal or correction of model outputs \cite{amershi2019guidelines}, AI-resilient design foregrounds two prerequisites: users must be able to \emph{notice} when the AI has made a consequential choice, and they must have sufficient \emph{context to judge} whether that choice is acceptable given their goals, preferences, and values. Without adequate support for these processes, users risk overlooking subtle but critical mistakes, or relying on model outputs without sufficient grounds for trust. Even in routine tasks such as search or summarization, unnoticed AI errors or contextually inappropriate outputs can mislead users and undermine decision-making~\cite{glassman2024ai}. 

This study takes the concept of AI resiliency into the legal domain, where the consequences of unnoticed or misjudged AI choices are particularly high. Legal professionals often rely on dense textual evidence, precedent, and careful interpretation. Even minor omissions or subtle rephrasing of the original text can have significant implications in this high-stakes professional context, like the difference between \texttt{will} versus \texttt{shall}. By presenting participants with prototypes with key AI-resilient interactions, i.e., only novel retrieval and rendering of existing text without generating new text, we investigated domain-specific challenges and design opportunities for designing tools that help legal experts remain in control of their reasoning and decision-making processes.


\subsection{Prototypes Demonstrated}\label{sec:prototypesbackground}
\subsubsection{Grammar-Preserving Text Saliency Modulation (GP-TSM)}
\label{sec:gptsm}

Grammar-Preserving Text Saliency Modulation (GP-TSM)~\cite{gu2024ai} is a recently developed text rendering technique that offers an AI-resilient alternative to traditional summarization. Instead of producing lossy summaries that may omit, misrepresent, or hallucinate information, GP-TSM recursively compresses sentences while preserving grammaticality, visually modulating word opacity to indicate levels of detail. This design allows readers to skim efficiently while retaining access to the full text, enabling them ``to notice, judge, and recover from automated [choices] they disagree with''~\cite{gu2024ai}. In our study, we used GP-TSM as one of the prototype tools to investigate how professionals might perceive value in AI-resilient reading tools that make reading and skimming more efficient while preserving the integrity of the original legal text, especially given the well-publicized risks of AI-generated text of any type which are already seeping into legal practice~\cite{nielsen2025law}.

\subsubsection{Abstract Explorer}
\label{sec:abstract_explorer}

Informed by Structure-Mapping Theory\footnote{In cognitive science, Structural Mapping Theory (SMT) is a well-established framework that describes how humans make comparisons by constructing mental representations that can be aligned in terms of their structural relations.}~\cite{gentner97,gentner2017}, AbstractExplorer~\cite{abstractexplorer} supports comparative close reading across large collections of scientific abstracts. Instead of abstracting away content through lossy representations, it leverages role-based highlighting, structural alignment, and ordering of original sentences to surface cross-document commonalities and differences, enabling users to skim and compare at scale while retaining author-written detail. 
In this work, we used AbstractExplorer as a design probe to investigate how professionals might perceive value in tools that allow them to notice and closely compare similar but distinct writing choices within analogous portions of documents. 


\subsubsection{CorpusStudio}
\label{sec: corpus_studio}
\emph{CorpusStudio}~\cite{dang2025corpusstudio} is an interactive writing environment that surfaces emergent community-specific writing patterns (norms) from a corpus of previous documents as the user writes. Developed initially for the HCI community, the corpora includes previously published systems based HCI papers. It surfaces emergent norms at the document level and at the sentence level. At the document level, it shows an ordered distribution of top-level section titles, giving users an view of the distribution over document structures in the corpus so that they can choose the structure of their own document in light of those precedents. 
At the sentence level, while users are writing, they can simultaneously retrieve as many as fifty contextually similar sentences\footnote{Contextually similar sentences have a combination of (1) a similar semantic embedding to the sentence or sentence fragment that the writer's cursor is currently next to and (2) occur in a section \textit{with a title} that has a similar semantic embedding to the section the writer's cursor is currently in.} from papers in the corpus, rendered with cognitive support to help emergent patterns and anomalies 'pop out'.
The prototype also features a tooltip that displays local context from the source papers where the sentences originated; bookmarks and notes allow writers to save sentences and capture insights; and anti-plagiarism measures, such as disabling copy and paste and displaying many examples in parallel to compare and abstract from (as predicted by Structure Mapping Theory~\cite{gentner2017}) rather than anchoring, even subconsciously, on one. 



In this work, we presented CorpusStudio as an exploratory prototype, inviting legal professionals to evaluate its fit as a writing support tool within their existing workflows. We chose CorpusStudio as an example writing prototype because legal writing is often precedent and template driven, and the prototype's corpora-based examples assist with comparison and also support making community writing norms and variation explicit, which aligns with how legal professionals learn from previously drafted examples.

Each of the demonstrated prototypes was shared by its respective developers, who provided permission for it to be evaluated in a new context.

\section{Related Work}
\subsection{AI and Law}

Research on AI and law spans decades, from early expert systems to current explorations of LLM-based tools. In addition to this legal scholarship, HCI researchers have increasingly studied how legal professionals and communities interact with technology, developing systems, surfacing risks, and examining design opportunities. In this section, we synthesize both traditions. We first review general concepts and use cases, then survey systems and tools, and finally highlight concerns that motivate our focus on better tooling support. This interdisciplinary view connects doctrinal and technical debates in AI \& Law with empirical HCI insights into usability, trust, and professional practice.

\subsubsection{AI and Law -- General Concepts, Use Cases}

A large number of prior research has mapped a wide spectrum of AI applications to legal work, from information retrieval and contract review to analytics, drafting, and decision-making support. Others trace how legal workflows have shifted from expert systems to data-driven and LLM-based approaches. The most prominent use of AI is arguably large language models (LLMs) for legal NLP tasks including information segmentation, retrieval, summarization, and extraction~\cite{siino2025exploring}. Historical retrospectives locate today’s methods within decades of developments in law, documenting the evolution from symbolic reasoning and argumentation to more computation-focused approaches based on machine learning and large corpora~\cite{bench2012history,ashley2019brief}. Regarding this evolution, prior work has framed ``legal analytics'' as a bridge between computational techniques and practice needs, emphasizing how text analysis, predictive modeling, and visualization can augment research and advisory workflows~\cite{ashley2017artificial}. Domain syntheses extend these mappings to specific practice areas such as criminal law and procedural contexts, highlighting applications such as risk assessment and evidence analysis alongside emerging issues such as algorithmic opacity and deepfakes~\cite{gouri2025ai}. In public and international law, scholars surface implications for personhood, autonomy, weapons systems, and governance design---areas where doctrinal adaptation is pressing~\cite{burri2018international}.

Prior work on practice and legal education emphasize operational readiness: how AI will affect firm economics and workflows, and how curricula should cultivate ``AI-literate'' legal reasoning~\cite{alarie2018artificial,ashley2012teaching,oskamp2002ai}. Together, these works position AI as a family of capabilities attachable to many legal tasks, while underscoring the need for interfaces that make automated choices legible and auditable. Our study advances this agenda by focusing on \emph{how} legal professionals actually read, compare, and draft dense legal texts. By introducing interview participants to recent novel prototypes that preserve original language and structure, we connect system capability surveys to situated use and identify design requirements for AI-powered systems that support high-stakes reading and writing.

\subsubsection{Tools for AI and Law}

A rich tradition of system development in AI \& Law has explored concrete technical approaches and interaction paradigms. Classic work on expert systems framed law as a domain for encoding jurisprudential reasoning, advancing early rule-based approaches to case analysis and legal decision support \cite{susskind1986expert}. Argumentation-based systems and logical formalisms have similarly been central, modeling how claims, premises, and rules interact in legal discourse \cite{bench1997argument,prakken2005ai}. These foundations established representational and reasoning infrastructures that remain influential in contemporary debates about the balance between symbolic and statistical methods.

Recent research has emphasized interface and workflow design. \citet{nielsen2025law} argue that law is particularly vulnerable to subtle AI influence and show how interface choices can mitigate risks by foregrounding transparency and user judgment. \textsc{LegisFlow}~\cite{legisflow} integrates temporal-aware LLM assistance into statutory research workflows, demonstrating opportunities for augmenting legal analysis while preserving access to primary materials. Other HCI-inspired systems take similar approaches: \citet{10.1145/3706599.3720167} designed a human–AI precedent search tool for Chinese law, emphasizing user control over intermediate steps and surfacing design challenges in collaborative research workflows. \citet{10.1145/3613904.3642743} evaluated \textsc{LegalWriter}, a structured writing support system for novice law students, showing how intelligent feedback improved the persuasiveness and clarity of case writing. \citet{10.1145/3176349.3176873} redesigned a document viewer for due diligence to better align with the workflows of legal professionals reviewing contracts. Together, these examples illustrate how HCI and AI \& Law communities converge on building systems that foreground usability, collaboration, and transparency.

Other perspectives call for renewed attention to ``good old-fashioned AI'' methods that leverage explicit rules and structured data when appropriate~\cite{bench2021need}, and for design philosophies that embed AI within broader professional values such as justice and the rule of law~\cite{greenstein2022preserving,lobel2023law}. Collectively, these systems show the breadth of tools, from expert systems to generative AI prototypes and HCI-inspired interfaces, being considered for legal practice. Yet, most share a common limitation: they optimize for task automation or abstract reasoning, rather than for the day-to-day reading and writing practices that define much of legal work. Our study builds on this lineage by treating systems like GP-TSM and AbstractExplorer not as end-products but as design probes. We hope this reframing surfaces concrete interface requirements that extend beyond what current AI \& Law tools explicitly support.

\subsubsection{Concerns about AI and Law: Why Better Tooling Support is Needed} 

While many works emphasize the potential of AI to enhance legal practice, others highlight deep concerns about fairness, accountability, and professional integrity. First, AI-powered legal interfaces can subtly bias judgments~\cite{nielsen2025law} or cause anxieties~\cite{solovey2025interacting} if design does not actively support user autonomy. Legal professionals’ behaviors can also shift as they become accustomed to AI-enabled tools, sometimes in ways that subtly reshape practice and expectations~\cite{10.1145/3343413.3378014}. Additionally, empirical and conceptual studies show that legal actors and lay people alike may struggle to \emph{evaluate} AI-generated outputs. Explainability features, such as attention-based highlights in legal summarization, can improve trust and efficiency, but not all explainability methods are equally effective~\cite{10.1145/3411763.3443441}. Even when users can tell human and machine-authored advice apart, they often still defer to LLM suggestions~\cite{schneiders2025objection}. \citet{10.1145/3581641.3584058} further show how trust in AI advice develops over repeated legal decision-making tasks, revealing that accuracy strongly shapes trust trajectories.  

HCI researchers also expose the sociotechnical challenges of legal AI. First, the disciplinary divides between computer science and law often lead to systematic errors and distortions when statutes are ``translated'' into software~\cite{10.1145/3686937}. \citet{10.1145/3512898} analyze the TREC Legal Track as a case of participatory design, showing how co-design with legal experts can help bridge this gap but also requires sustained structures for meaningful engagement. Studies of applied contexts highlight risks at the margins. For instance, \citet{10.1145/3706599.3720532} show that legal professionals face major challenges in handling technology-enabled abuse in intimate partner violence cases, revealing gaps in education and tooling, while \citet{10.1145/3613905.3637108} analyze the design of an eviction defense web tool, underscoring tensions between expert authority and lay participation. Broader civic perspectives also show the stakes: \citet{10.1145/3706598.3714322} bring cross-national lay perspectives into debates on generative AI’s future, underscoring the societal and democratic dimensions of legal technology adoption.  

Scholars underscore normative risks as well. \citet{wachter2021fairness} argue that fairness cannot be automated, pointing to gaps between EU non-discrimination law and algorithmic decision-making. \citet{davis2018law,davis2019artificial} critiques the notion of ``mindless'' AI legal reasoning and explores the limits of attributing wisdom or jurisprudential authority to machines. Ethical surveys and doctrinal analyses further highlight unresolved tensions around explainability, transparency, and accountability in AI-supported legal decision-making \cite{richmond2024explainable,gordon2021ai,surden2020ethics}. 

These critiques converge on a central concern: without careful design, AI can erode professional judgment and undermine the rule of law \cite{greenstein2022preserving}.
Taken together, these concerns motivate our focus on \emph{AI-resilient} interfaces. Rather than assuming correctness or fit, we investigate how tools can be designed to safeguard accuracy, accountability, and human control in high-stakes legal work.

\section{Interview Methodology }
We conducted a semi-structured interview study with legal professionals to learn about their approaches to reading and writing documents, 
and understand their expectations for tools that support their workflow through the lens of three atypical AI-powered reading and writing support tools described in \Cref{sec:prototypesbackground}. While these prototypes were originally designed and evaluated in the HCI community, we invited participants from the legal domain to view the systems through the lens of their own context and to conduct a contextual fit evaluation (whether they would use the tools, for which document types, how they would use them, and at what stages in the reading or writing process) that revealed as much or more about their own contexts as it did about the particular prototypes. Participants brought a broad range of experience that spanned many areas of law. Each session lasted 60 to 75 minutes and was conducted remotely via Zoom.

\subsection{Participants}
We recruited twenty-two participants via academic and legal professional networks, with representation spanning multiple countries and continents. All participants were over 18 years of age and fluent in English. The participants had an average of 16.5 years in the legal profession, with 18.2 years reading and 16.7 years writing legal documents (\cref{tab:participant_info}). Roles spanned the legal ecosystem and across sectors (\cref{tab: participant_role}). Reading and writing were central to participants’ work, with frequent reliance on skimming and engagement in using prior documents (\cref{tab:freq_behaviors}).  

\begin{table}[]
\centering
\newcolumntype{L}{>{\raggedright\arraybackslash}X}
\newcolumntype{P}[1]{>{\raggedright\arraybackslash}p{#1}}
\renewcommand{\arraystretch}{1.2}
\setlength{\tabcolsep}{4pt}
\begin{tabularx}{\linewidth}{P{2cm}P{3cm}L}
\toprule
\textbf{Category} & \textbf{Details} &  \\ 
\midrule
Participants & \multicolumn{2}{l}{Total: 22 (11 men, 11 women)} \\
& \multicolumn{2}{l}{Age: 18-24 (1 participant), 25-34 (5), 35-44 (8), 45-54 (3), 54+ (5)} \\
& \multicolumn{2}{P{6cm}}{Criteria: Over 18 years old, fluent in English, legal professionals involved in writing or reviewing legal documents} \bigskip\\
\multirow{2}{2cm}{Experience} & Years in profession & 16.5 years \\ & Reading legal documents & 18.2 years \\
& Writing legal documents  & 16.7 years \bigskip\\
\bottomrule
\end{tabularx}
\caption{Participant information}
\label{tab:participant_info}
\end{table}

\begin{table}[]
\centering
\newcolumntype{L}{>{\raggedright\arraybackslash}X}
\newcolumntype{P}[1]{>{\raggedright\arraybackslash}p{#1}}
\renewcommand{\arraystretch}{1.2}
\setlength{\tabcolsep}{4pt}
\begin{tabularx}{\linewidth}{P{3cm}P{1.5cm}L}
\toprule
\textbf{Category} & \textbf{Count} & \textbf{Included Titles} \\ 
\midrule
General Counsel & 3 & General Counsel (2), General Counsel and Chief Agent (1) \\
Lawyer / Attorney & 3 & Lawyer (2), Attorney (1) \\
Legal Academic & 4 & Lecturer-in-Law, Lecturer on Law (Legal Writing), Law Professor (Lawyer before that), Visiting Assistant Professor of Law \\
Legal Support / Operations & 2 & Legal Operations Manager (1), Knowledge Management Intern/Law Clerk (1) \\
Associates & 2 & Associate (1), Senior Associate (1) \\
CEO Roles & 2 & CEO \& Principal (1), CEO-retired (1) \\
Public Service / Political & 1 & Formerly Pro Bono Counsel (Current judicial candidate) \\
Advocacy / Advisory / Design & 3 & Advocate (1), Advisor (1), Program Development Coordinator/Designer (1) \\
Retired (unspecified) & 1 & Retired \\
\midrule
\textbf{Total} & \textbf{22} &  \\
\bottomrule
\end{tabularx}
\caption{Roles of Participants}
\label{tab: participant_role}
\end{table}




\begin{table}[h!]
\centering
\begin{tabular}{llcc}
\toprule
\textbf{Activity} & \textbf{Frequency} & \textbf{n} & \textbf{\%} \\
\midrule
\multirow{3}{*}{Reading legal documents} 
  & Few times/day       & 11 & 50.0 \\
  & Few times/week      & 9  & 40.9 \\
  & Few times/month     & 2  & 9.1 \\
\midrule
\multirow{4}{*}{Writing or revising legal documents} 
  & Few times/day       & 8  & 36.4 \\
  & Few times/week      & 8  & 36.4 \\
  & Few times/month     & 4  & 18.2 \\
  & Rarely              & 2  & 9.1 \\
\midrule
\multirow{4}{*}{Rely on skimming when reading long documents} 
  & Very often          & 5  & 22.7 \\
  & Often               & 10 & 45.5 \\
  & Occasionally        & 6  & 27.3 \\
  & Rarely              & 1  & 4.5 \\
\midrule
\multirow{4}{*}{Compare their work to prior related documents when writing} 
  & Very often          & 8  & 36.4 \\
  & Often               & 9  & 40.9 \\
  & Occasionally        & 2  & 9.1 \\
  & Rarely              & 3  & 13.6 \\
\bottomrule
\end{tabular}
\vspace{0.5em}
\caption{Frequency of reading, writing, and document review and drafting behaviors (N=22).}
\label{tab:freq_behaviors}
\end{table}

Most participants routinely used support tools for reading and writing \cref{tab:tool_usage}. Common tool types included document editors (Microsoft Word/Google Docs; 22/22), commercial AI assistants (e.g. ChatGPT/Claude/Copilot; 20/22), in-house or self made AI tools (1/22 each), legal/clinical research platforms (e.g. Westlaw/LexisNexis; 17/22), writing-enhancement tools (e.g. Grammarly/Wordtune; 5/22), and occasional use of collaborative design/diagramming tools (Figma, Miro; 1/22 each). We examine how participants use these tools in \cref{sec: results_read_write_practices}.

\begin{table}[h!]
\centering
\begin{tabular}{lcc}
\toprule
\textbf{Frequency} & \textbf{Writing Support Tools} & \textbf{Reading Support Tools} \\
\midrule
Once a day or more      & 7 (31.8\%)  & 6 (27.3\%) \\
Few times a week        & 8 (36.4\%)  & 3 (13.6\%) \\
Few times a month       & 3 (13.6\%)  & 5 (22.7\%) \\
Only a few times        & 4 (18.2\%)  & 6 (27.3\%) \\
Never                   & --          & 2 (9.1\%)  \\
\bottomrule
\end{tabular}
\vspace{0.5em}
\caption{Frequency of writing and reading support tool usage among participants (N=22).}
\label{tab:tool_usage}
\end{table}


\subsection{Interview Protocol}
\subsubsection{Consent and Pre-Study Survey}
After giving informed consent, each participant completed a brief survey about their demographics, years of experience in the legal profession, experience with reading and writing legal documents, use and experience with reading and writing support tools in their workflow support tools, tools currently used, and questions relevant to their experience in their profession. All survey questions are included in \cref{app:pre-study}.

\subsubsection{Background and Current Practices}
Before introducing the tools, we first learned about each participant’s professional context (background and current role). We then explored reading practices: types of legal documents they typically read, what they considered challenging about reading these documents, and how they approached reading these documents. We repeated the same structure of questions for writing (document types, challenges, and typical approach/strategies). The exact questions can be found in \cref{app:bacp-questions}.

\subsubsection{Tool Exploration}
Participants were shown three prototype tools, i.e., the two designed for reading support, GP-TSM (~\cref{sec:gptsm}) and AbstractExplorer~(\cref{sec:abstract_explorer}), and one for writing support, CorpusStudio~(\cref{sec: corpus_studio}). We presented a live demo of each tool, followed by a discussion about the participant's feedback on it in the context of their profession or current role. Specifically, the conversation for each tool was structured around three themes: first, initial impressions of the tool; second, target materials, i.e., what they would want to skim or read at scale for GP-TSM and AbstractExplorer, and what corpora they would want to query when drafting with CorpusStudio; and third, adjacent needs, i.e., given the tool, what additional capabilities would be useful (within their context for GP-TSM, for exploring document collections with AbstractExplorer, and for indexing into specific sections of long documents with CorpusStudio). For each tool, we concluded with an unstructured portion for participants to surface any topics not covered by our questions. The questions can be found in \Cref{app: tool-qs}. Note that for each tool, we informed participants that the current prototypes were originally designed for users in the HCI community, and therefore the example content in each tool was from HCI venues rather than legal documents. We explained that our goal was to gather feedback from legal professionals to inform future tool law-specific design. Given that this tool exists, we asked, “what type of tools would be useful within your context?” We framed the question this way to reduce social desirability bias because a tool was already presented ~\cite{grimm2010social}. The order of tool interaction was counterbalanced, with some participants viewing the reading tools first and others starting with the writing tool.


\paragraph{Protocol modification.}
In the first session (P1), after a brief walkthrough demo, the participant was asked to directly interact with each prototype (e.g., retrieving sentences in CorpusStudio). Because the example corpora were drawn from HCI rather than law, P1’s ad-hoc queries returned results that were not legally meaningful; both our observations and P1’s feedback indicated that extended hands-on time added limited value. Beginning with P2, we therefore replaced the open interaction period with an extended discussion on contextual-fit focused on whether/when/how the tools could be useful in legal work. This better aligned with our focus on interview responses rather than tool interaction metrics. All other aspects of the protocol remained the same.



\subsubsection{Open-Ended Questions} 
We concluded the study with a set of open-ended questions to gather participants’ reflections on current and future tool use. Participants were asked to describe the pros and cons of either AI-based or non-AI-based tools they currently use for writing and reading. They were also invited to reflect on tasks they would not want AI to perform, even if those tasks were complex, as well as the factors they consider before adopting new tools. Finally, participants were asked whether they had any risks or sensitivities they would be concerned about when using such tools in their work. All open-ended questions are included in \cref{app:open-questions}.
\section{Results}
The two lead co-authors jointly led all interviews and analyzed the data using a \textit{thematic analysis} approach \cite{terry2017thematic}. Our data included responses to the pre-study survey, notes taken during the demos and discussions, and verbatim transcripts. The first author synthesized the survey results and discussed them with the second author. Both leads independently reviewed every transcript, then conducted deeper analyses organized by tool and by open-ended question to surface emergent themes, insights, outliers, and suggestions. Throughout the analysis, co-authors discussed results, compared interpretations, reconciled differences or overlapping themes, and iteratively refined the results presented in this section. All co-authors reviewed and approved the final results report. 

\subsection{Reading and Writing in Legal Practice: Insights for System Design}
\label{sec: results_read_write_practices}
To understand the reading and writing practices of legal professionals, we explored the types of documents they engage with, the challenges they encounter, and their typical approaches to these tasks. These insights provided context for interpreting their feedback and helped us identify opportunities for tool and interface improvements or developments. Below, we present the key themes that emerged from their responses.

\paragraph{Documents Commonly Read and Written} Participants reported engaging with a diverse range of legal texts in both reading and writing tasks, spanning formal sources like case law, statutes, and litigation materials to contracts, internal communications, and academic or policy documents. Their written outputs reflected similar variety, covering litigation filings, transactional contracts, advisory memos, and scholarly works. This breadth underscores the complexity and scope of legal workflows across roles. For detailed descriptions and participant references, see ~\cref{app:docs-read-written}.

\paragraph{Reading was often described as laborious, with participants citing both volume and density as persistent challenges (P1, P3, P5, P12, P13, P14, P16, P17, P18, P19, P20, P21).} Some documents contained hundreds or thousands of pages, and contracts and statutes were particularly dense and verbose. Participants reported mental fatigue from parsing long linear text, and reading was time consuming, particularly searching for small snippets of information buried in lengthy documents (P1, P10, P12). Many also described the difficulty of extracting relevant information (P1, P2, P5, P6, P12, P17, P21, P22). As P6 described, the challenge was often less about decoding and interpreting the text and more about “extracting what’s most relevant,” due to the abundance of unnecessary information. 

Understanding terminology and jargon was another hurdle (P4, P5, P7, P9, P11, P17, P22), with some pointing to inherent ambiguity and intentional obfuscation (P4), or legalese (P7). P22 mentioned the pedagogical difficulty of drawing attention to important terms in dry, complex language. When terms or interpretations were unclear, participants consulted colleagues or other legal exports (P7, P18) and relied on standardization to reduce ambiguity (P7). 

Structural complexity added to the difficulty (P9, P12, P20). Documents varied in layout, statutes and regulations, for example, follow nested, code-like structures rather than standard expository formats (P20). Cross-referencing between documents or laws further compounded the challenge (P4, P5, P7, P13, P15), with meaning often buried in obscure references or dep within a document (P4). Poor navigation tools, such as ineffective tables of contents, also hindered reading (P12) 

Writing quality also impacted readability. Participants cited poor or inconsistent writing quality (P13, P14, P20, P21), or texts that failed to serve the reader's needs (P20, P21). Two noted the pressure of limited time (P11, P17), especially when needing to operate "at the speed of business" (P11). A recurring concern was the risks of skimming (P9, P10, P16, P18), which participants said could lead to serious mistakes, either by missing hidden clauses (P16), glossing over critical details (P18), or relying on assumptions from prior experience (P9). As P10 noted, “skimming is an invitation to [make] mistakes,”

Language-specific issues further complicated reading. One participant highlighted the difficulty of working across multiple languages and ensuring accurate interpretation (P21). Another noted that while language reuse is common and accepted, it makes comparison labor intensive and prone to error (P22). 

Finally, two participants reported AI-related reaching challenges (P13, P19). These included inaccurate or outdated interpretations (P13), ineffective keyword searches due to formatting issues, and privacy limitations restricting use of tools like OpenAI (P19).

\paragraph{When faced with a complex and often overwhelming amount of information, participants relied on a combination of digital tools and strategies shaped by their professional judgment and experience to manage their reading tasks.} They used AI-enabled legal databases such as LexisNexis, Juta, SAFLII, and Westlaw (P1, P8, P12, P15, P16, P18, P21), entering specific legal questions to retrieve and skim top results for relevance.

Some also used general-purpose AI tools to read summaries (P2, P13, P15), answer questions about a document (P8, P11, P12, P21), or focus on key sections (P16, P18). P8 prompted ChatGPT or Claude “as if [to] a senior attorney” to flag issues while relying on personal expertise to fill gaps. P11 used NotebookLM Pro and Paxton for risk-assessments. P12 queried ChatGPT, estimating a 30/30/30 split between helpful, irrelevant, and unexpectedly insightful responses (e.g., learning new or modernized terminology they then verified via Google). Others prompted AI tools to surface possibly intentional hidden details (P16), focus the AI model on pre-identified problematic passages to compare findings with their own understanding while avoiding full-document summarization to prevent losing critical detail (P18), or extract specific answers through custom prompts (P21).

Participants used summarization tools at different stages. Some used them as a first read to generate an overview of the document (P11, P16, P17, P21). P16 said, “AI has taken over a huge portion of my reading... Harvey AI is my favorite tool... it has enhanced features for analysis, summarizing, and rewording in the voice that you need.” P17 used ChatGPT and CoPilot to summarize text and flag risky or important language.  P21 outlined documents with AI tools, later validating the output manually. Others used tools such as CoPilot or in-house models, tailored to their specific legal domain or organization, after a manual read through to confirm their understanding (P9, P15, P19). Generative tools like Harvey or Agora were used to validate interpretations, but only after an initial close read.

Traditional methods remained common. Optical character recognition (OCR) enabled text search or text-to-speech for multitasking reviews (P1, P11, P15). Control-F was widely used: “I love Control-F… find what it is that I need, and then once I do, I’ll read all around it, and then I’ll scroll for context” (P3; also used by P8, P12, P15, P17, P19, P22). Participants evaluated edge cases (P3, P4), redlined changes (P4, P8, P13), and used familiarity with structure, described as “muscle memory,” to quickly locate information (P4, P5). P5 noted, “I know how much my finger needs to move on the mouse to get to the indemnification clause." Cross-document helped build context and align precedent (P5, P9, P19). 


Reading was described as iterative and purpose driven. Some shifted from deep reads to strategic skims followed by targeted deep dives (P2, P21). Many skimmed first, then revisited key sections (P3, P6, P7, P9, P11, P14, P17, P19, P20, P21, P22), adjusting based on task (e.g., decision making vs. detailed feedback) (P14). Strategies varied by document type. For example, skimming headers in complaints or skipping introductions in law review articles to reach core analysis (P20). Some read multiple times (P10, P18, P19), starting with a full read before narrowing to decisive passages (P10, P19). Reading depth also depended on content importance, sender, and audience (e.g., client-facing vs. internal memo), shaping how closely a document was reviewed (P19, P21).

\paragraph{Participants described one challenge of writing as it being a balancing act between producing clear, concise documents without becoming bulky or overly argumentative (P1, P4, P13, P20) while ensuring legal accuracy (P6, P7, P9, P12, P13) and accessibility (P7, P20; make it understandable without omitting critical information).} Adhering to expected norms and tailoring documents to audience expectations was one of the most commonly reported challenges (P2, P3, P7, P11, P12, P13, P16, P18, P21, P22). Specifically, deciding how much detail to include, anticipating follow-up responses or questions, matching expected structure, style, and language, and ensuring nothing is assumed so that a reader who may not share the same context or knowledge could still follow. Writers also wrestled with cross-referencing across documents and laws (P1, P13), maintaining consistency of style and structure (P2, P4, P5, P11, P15) and adhering to proper standards, style, tone and strict formatting that could result in rejection risk for deviations (P5, P6, P8,  P17, P19, P22). Another common challenge was translating complex material while preserving compliance and clarity (P7, P9, P12, P13, P20, P22).
Several practical challenges further complicated the legal drafting process. These included limitations in traditional drafting tools like Word and PDF (P4), inconsistencies in version control across teams and jurisdictions (P15), the time burden of drafting and revisions (P11), the need to stay aligned with evolving industry regulations (P16), and the manual overhead of incorporating amendments into existing documents (P13). 

The blank page problem was also mentioned; participants noted that templates helped, but could be potentially limiting and prevent one from spotting omissions (P10, P15). One participant noted that drafting feedback that is both supportive and trenchant could be challenging as it required extra effort (P14). 
Existing AI support tools brought a mix of benefits and challenges: these tools, even legal-specific systems, struggled with translation tasks (P9), often produced wordy or overly legalistic text requiring simplification (P11), could be inexact enough that revising them took as much time or effort as manual drafting (P15), and introduced review complexity when aligning human and AI edits to preserve tone, audience fit, and substance (P19). Some constraints were role or venue specific, including the diminished fluency that comes from infrequent full-length drafting (P19), and academic norms that limit the reuse or extensive quoting of clear, well established prior descriptions, requiring restatement in original language (P22).

\paragraph{Turning to participants’ writing approaches, some reported engaging in a front-loaded planning phase prior to drafting during their writing process.} Several began with outlines, utilizing different methods: note taking while reading (with key points and page numbers), mind maps to organize ideas, brain dumps in Google Docs to create high-level outlines (P1, P11, P13, P20, P21). Many clarified the communication objective (P3, P21), and considered audience expectations by venue (P3, P12, P13, P18, P22): academics expect deep theory and citations; regulators value concise summaries with optional detail; businesses prioritize actionable steps with legal and commercial justification (P13). 

Reusable scaffolds and collaboration were central. Participants routinely referenced industry standard templates, internal precedent databases, and similar prior documents, even when there is no exact match they would adapt from nearby examples (P4, P5, P7, P8, P9, P10, P11, P15, P16, P17, P19, P20, P22). 

Participants also tailored scope by deciding how much detail is necessary (P3, P13, P18), for example by focusing on a top three message set, asking what they would regret not including, and avoiding fascinating but irrelevant details (P13). One participant applied content strategy principles to enhance navigability, simplifying to essentials and using accessibility best practices in color, layout, and visual clarity (P7). 

They incorporate client and colleague input throughout (P5, P6, P7, P22). For example, some rely heavily on colleague feedback: P6 noted, “I want to do it more autonomously, but I need validation currently.” Others involve stakeholder/user research (P7), and iterate across multiple rounds of peer review (P22). Consistent with this collaborative approach, P22’s approach to writing was to focus on issues that readers raise and let feedback guide development.

Generative AI tools served as support aids that accelerated the drafting process, followed by manual expert revision (P5, P9, P11, P13, P15, P16, P17, P18, P19, P21). Participants use legal-AI tools like Harvey and Paxton to locate templates and produce rough drafts that were then refined manually. First drafts from these tools were verified, cleaned, and expert approved. When needed, they instructed these tools to draft documents from scratch with explicit context and instructions such as jurisdiction of interest and desired clauses. They would use playbooks or previous documents with tools like NotebookLM as sources to guide tone and content (P11). 

Some preferred LLMs for search and aggregation rather than generation (P15) while others used generative AI for generating text and suggestions `around' the main writing task. For example, P16 used Harvey, Copilot, and ChatGPT to improve writing quality, double check for errors or outdated language, and integrate up-to-date regulatory references; they noted that AI “has made the quality of my outputs very much higher” and is “the only way you have, as far as I know, to double check your work with no bias” (P16). P17 drafts documents in their own style first, then uses ChatGPT or Copilot to reformat to required instructions while ensuring to redact personal identifiable information for confidentiality. Iteration is the norm across these steps. P17 found AI-based tools to be time saving, allowing for greater project capacity, though they always review outputs due to possible hallucinations or incorrect reasoning. P19 used AI-based tools to help with formatting, grammar, tone, and persuasion, but not for substance or legal drafting; they maintained external prompt libraries to control prompt security and avoid platform retention of information. Similarly, other participants also used these tools for grammar checks (P16, P20). Some participants used custom built AI tools that were either compliant with regulation or personalized with their profile, clients, and legal role (P18, P21); despite using custom built tools, both participants still reviewed, revised, and verified the generated information to ensure accuracy and relevance.


\subsection{Reading Tool 1: Grammar-Preserving Text Saliency Modulation (GP-TSM)}
\paragraph{Understanding How the Tool Works}
Initial reactions to \emph{GP-TSM} for some participants centered on transparency of the system design (P1, P7, P14). They wanted to understand the system’s decision logic: what gets de-emphasized, what is kept, and why. Several wanted transparency into the decision process, including which words were down-weighted and insights/rationale for the model that was used (P7), and how we trained the model on the grammar-preserving aspect to determine what matters (P14). Others were drawn to the functionality itself (P6, P9): the idea of emphasizing salient text and graying the rest was “intriguing,” with the caveat that it might take some adjustment to get used to, especially on longer documents. As P6 stated, “I like the idea of emphasizing words. At first glance, it's kind of it feels like it would be kind of hard to read for longer documents, but I think it's more of a thing you get used to.”

\paragraph{Speed \& Orientation Payoff}
Time savings and efficiency from being able to quickly skim and re-read emerged as a recurring benefit of using \emph{GP-TSM}. Participants anticipated using GP-TSM to “speed read” or accelerate first-pass comprehension (P1, P5, P10, P12, P15, P16, P19), including filtering “fluff” (P5), getting oriented to unfamiliar material before a deep read (P1, P4), or producing a faithful “TLDR version" within a longer document while retaining full fidelity to the original source (P9). As P1 explained, “most of my life is based on skimming documents... I think it's quite useful to have a tool like that,” emphasizing how well GP-TSM fit into their existing reading habits. Multiple participants described the rendering as making important ideas “pop,” feel “productive... supporting faster comprehension” and support fluid skimming with the option to still doing a more thorough reading when needed (P1, P5, P19, P20). P12 described it as efficient, rapid, and notable for its usefulness in cutting through dense text and they saw value in using it to grasp broad strokes quickly. One participant summarized the benefit for them succinctly: “this is basically sparing me from doing the first reading” (P16). 

\paragraph{Mimics Current Practices}
Several participants (P3, P8, P10, P12, P14, P16, P18, P21) noted that \emph{GP-TSM's} use of black and varying shades of gray to emphasize or visually de-emphasize elements of text closely mirrors their own highlighting practices, which they find effective for supporting memory recall during review. P8 does this in Word with Control+F and highlighting; P12 compared it favorably to “old-school” circling of keywords to stay alert; and P14 stated, “... it looks like a rendered version of what I think most readers do, right, except that you're making it easier. Like I don't have to make those calls myself... you've made those calls.” P16 appreciated that the tool automatically surfaces important content, effectively doing the underlining they normally perform on a second pass, and P18 framed it as everyday practice (“It’s what we do every day without having this help… This could be a very, very nice thing”) and as a precursor to mind-mapping (“It’s the first thing you do before making balloons [mind maps], because it gives you a visual emergence of what's important”). They also valued the underlying model’s ability to simulate the thought process of tailoring detail to the audience, something they already do manually. Though this was a recurring theme among many participants, not everyone saw this as aligning with their practices: P4 noted, “When I’m doing real work... it’s not how I read,” adding that grammatical preservation, while technically correct, could still interrupt their reading flow.

\paragraph{Caution with Initial Reads and Fear of Missing Critical Details}
Participants raised concerns about both the use of \emph{GP-TSM} for first reads and the potential for missing critical information. One participant stated, “I wouldn't want to use it, at least initially, without a lot of trust… I wouldn't use it for my first read of anything,” preferring instead to apply \emph{GP-TSM} after a full read to refresh memory and save time (P10). The central concern was missing important information due to de-emphasis, particularly modifiers, sentence structure, cross-references, and grammatically critical elements, where small words can change meaning (P2, P4, P5, P10, P11, P13, P14, P17, P21). P13 called it a “fine line” between helpful focus and loss of context, citing terms like “at least” or “at minimum,” and P21 emphasized obligations in legal practice to read the original text carefully, where distinctions such as “must” versus “may” matter. Similarly, P11 described \emph{GP-TSM} as “totally useless,” explaining that as a lawyer, they need to read every single word of a document, since nuance and small details (e.g., commas) can be critically important in legal analysis. In their view, de-emphasizing parts of a text is “the opposite of helpful.” P17 suggested that if the core meaning were preserved, concerns would lessen.

\paragraph{Familiarity and Customization}
A preference for familiarity and customization to support a more controlled reading experience was another theme that emerged. Participants asked for integration into tools they already use (e.g., Word, Adobe Reader) (P6) and for customization so emphasis or de-emphasis aligned with their specific needs (P1, P7, P8, P9, P10, P14, P22). This ranged from accessible color gradations (P7) to emphasizing only negotiable contract clauses (P8), an interactive way to restore and annotate de-emphasized text (P14), and flexible configuration options with all text preserved (P22). Several also wanted simple toggles to see only the emphasized text or collapse less important material and then expand as needed, with one participant likening this to navigating source code via function calls and definitions (P2, P4, P9, P22). Others asked to apply the technique to full documents (P12) and reiterated the need to “build trust” in the output over time (P22).

\paragraph{Readability and Visual Adaptation}
Participants expressed mixed reactions to the readability of GP-TSM's rendering. While some found it potentially straightforward and adoptable, others were more hesitant, noting concerns about initial readability due to how the text was rendered. P6 noted that the tool felt “kind of hard to read for longer documents” at first glance, though they believed it might become easier with use. Similarly, P5 stated that the interface “looked a little weird” initially, pointing out the different visual gradients in some sentences. P7 raised curiosity about how the tool would perform with legal documents that use nonlinear formatting, and expressed that they pause while reading because of the way text is segmented, even though it feels grammatically correct. P13 offered a generally positive view, describing the output as “pretty good,” but described a tension between their fast, detail-oriented reading style and the emphasis and de-emphasis of the tool, which caused them to slow down and question the selection of bold words, despite appreciating that the overall text remained coherent.

\paragraph{Use Cases for \emph{GP-TSM}: Documents to Skim or Read at Scale and Documents to Not}
Participants identified three main use cases where a skimming or summarization tool like GP-TSM could support legal reading: overly long or formulaic documents, formal or academic materials, and practice-facing communications. However, others cautioned against its use for high-stakes or citation-sensitive texts such as contracts, statutes, or litigation documents, or in contexts where nuance and precision are critical. These diverging views reflect differing perceptions of risk, context, and the tradeoff between efficiency and legal fidelity. See \cref{app:gptsm-uses} for full details.

\paragraph{Comparing GP-TSM Output with AI-Generated Summaries} 
Some participants compared \emph{GP-TSM} and AI-generated summaries. Among those, more participants preferred GP-TSM (P6, P9, P10, P14, P16, P17, P18, P19, P20, P22), while a smaller group preferred generated summaries (P8, P11, P15, P21). Participants who favored GP-TSM emphasized that keeping the full text in view allows "selective skimming" (P9, P10): they could move fluidly between the summary (emphasized text) and the underlying message without losing context, switching tools, or looking to reference the original document, making it easier to verify claims and dive into the full text when needed. As one participant put it, “this would probably be more useful... because I have the context that was hidden, kind of more ready to check. With the ChatGPT way, I’d have to look for the full context in the original document… While here... it’s more readily available to me to read the grayed-out text” (P6). Several participants connected this aspect of the tool to trust and accuracy: they perceived the lower (it is actually a lack of) risk of hallucinations (P9), found the saliency based rendering more trustworthy than black box summarization (P10), and appreciated seeing what was emphasized without “losing what was dropped.. That’s my big problem with the Gen AI summaries... I don’t know what got left out” (P14). Others highlighted control and workflow fit, as P16 stated, “I would trust me with this support much better than Harvey in reading a contract” and P18 noted, “This is better, actually, because we have more control. I like this more than AI summarizing” (P18). Participants also described GP-TSM as faster to use in practice (P19), helpful for maintaining authorial voice and intellectual integrity (e.g., avoiding copy-paste temptations they associated with summaries) (P19), more (actually, completely) grounded in the original language (P20), and preferable given skepticism that AI-generated summaries are “always wrong… even for non-nuanced content” and unreliable in both scholarship and legal contexts (P22). 

Although P19 preferred GP-TSM over AI-generated summaries, they noted that it does not easily support the sharing of excerpts, which could be a limitation in workflows that rely on quickly copying or distributing summarized content. A few participants described using GP-TSM in complementary ways. For example, using it for personal reading, comprehension, and notetaking (P16), or for skimming the full document to understand its overall structure and content and use tools like Harvey for summarizing information to share with others. P17 would use GP-TSM to skim and use tools like ChatGPT when they needed to locate specific information or extract particular points (P17).

The four participants who preferred AI-generated summaries emphasized factors like interpretive synthesis (P11) and ease for users who might otherwise avoid reading (P8). One argued that summaries “provide interpretation, not just modified original text,” can simulate the value of human discussion, and support comprehension for complex materials (P11). Another found GP-TSM’s de-emphasized text visually difficult to ignore, stating “I probably would trust a generated summary more… I actually find it hard to block out those, the lighter gray words… so this might be a cognitive behavioral economics type thing, but I just, I’m still trying to read the whole thing. It’s just more difficult” (P15). One participant asked, “why wouldn’t I just use an AI tool to generate a summary?” (P21) They regarded GP-TSM as less useful for experienced legal professionals (who already skim effectively), potentially more useful for non-lawyers or for low risk contexts, and felt GP-TSM’s importance overlay lacked the “clear cut” outputs provided by generated summaries (P21).

\paragraph{From the Source: Participants' Ideas for New Support Tools, Inspired by \emph{GP-TSM}}
Participants proposed feature enhancements for GP-TSM, as well as innovative reading support tools inspired by it, tailored to the specific needs of their roles within the legal profession.

One participant centered their ideas around making contracts and other legal records actionable rather than static (P2). They proposed “living contracts” that treat metadata (e.g., renewal dates) as triggers for alerts, workflows, and compliance tasks, shifting from passive storage to active monitoring and impact-aware tracking when terms or jurisdictions change. Rather than analytics in isolation, they argued for contract lifecycle management style features that connect analysis to concrete follow-ups (e.g., annual reporting reminders). 

Another set of proposals focused on precision retrieval and structure-aware navigation. Suggestions included diff-style reading that surfaces only what changed across versions or across contracts built from the same model templates, layered navigation that treats legal text like code (includes/references) so readers can traverse from surface concepts into deeper materials, and clause banks for reusable language (P4, P5). Participants also asked for sentence and section level “jump to” tools (e.g., clause summaries with direct links), filters to limit attention to specific provisions, toggles to show and hide de-emphasized text, and restoration/annotation controls for subtle phrases (P9, P14, P15). 

Another participant emphasized terminology centric search, such as synonyms, frequency, context, industry terms, and how terminology evolves across legislators, regulators, and enforcers, to support piecing together divergent perspectives (P12). Negotiation-oriented features included automatically flagging non-standard language and deviations from precedent, and relevance filters keyed to counter parties or matters (P8, P18). One participant suggested tooling support for handling citations and distinguishing from more versus less authoritative sources while navigating briefs and case law (P20). Another participant also emphasized communication and negotiation-based context awareness; they asked for an audience-aware writing assistant that adjusts tone and concision based on prior exchanges, a negotiation-aware contract tool that remembers past redlines or sticking points and flags terms previously accepted or contested, and a history-based reference that anticipates what a recipient will need (P3). Two participants were interested in further testing of GP-TSM to evaluate accessibility and understanding: they proposed a comprehension test comparing bolded only and full text (complete text without GPT-SM rendering) reading and stressing that tools must preserve granular detail, understand legal thresholds/terminology, and be trained for jurisdictional variation (P7, P13).  

Other areas of interest focused on reading experience, summarization, and adaptation to the reader. Ideas ranged from speed reading interfaces (narrow columns, automatic chunking into conceptual units, teleprompter-style pacing) to agentic systems that learn individual preferences and adjust the rendering accordingly (P10). One participant preferred interpretive, non-verbatim summaries, more like talking to a person, over in-place emphasis alone, citing tools that already deliver stepwise breakdowns and better follow-ups (P11). 

Others proposed client facing readability support that had readability grading, plain English (layperson) rewrites, translation, and simplification (P7, P15). One participant proposed a two phase division of labor: use GP-TSM to understand a contract personally, then a different tool to draft the outward facing summary for colleagues, underscoring the desire for distinct workflows for comprehension vs. communication (P16). Another envisioned concept mapping support that makes relationships visible for clearer thinking to support the next phase of more elegant writing (P18). 

Finally, participants stressed fit with everyday tools and smoother flow. Requests included add-ins for Word, Adobe Reader, or email to avoid copy-paste friction, overlays that keep readers “on the page” without tab switching, and simpler, prompt light workflows (e.g., click “Review,” drag files into the interface, and the interface auto populates standard questions, then “Go”) (P6, P19, P21). They also proposed clearer source reliability cues, like color coded indicators, to support judgment of AI outputs, along with reader controlled prioritization (e.g., “show me the facts” vs. “show me the law”), and highlighting that can surface verbosity during writing (P21, P22).

\subsection{Reading Tool 2: AbstractExplorer}
To explore how legal professionals might engage with novel cross-document comparison interfaces, we introduced them to \textit{AbstractExplorer}, a prototype originally developed for academic literature review. The tool organizes large collections of documents by highlighting and aligning analogous sentences by common shared structure, enabling comparative close reading at scale. Participants reflected on both the opportunities and challenges of adapting such an interface for legal contexts. 

\paragraph{Digesting and comparing large amounts of information.}
Many participants described \textit{AbstractExplorer} as an efficient and creative way to surface patterns across large sets of documents (P3, P5, P7, P9, P12, P14, P19). P19 appreciated the lateral reading process this tool enabled, noting that they could \textit{``review all the documents at once.''} P3 reflected that \textit{``it’s a creative way to look at how to digest information and see different patterns,''} while P5 emphasized the speed with which users could skim large corpora: \textit{``It’s incredible how fast you can just look through so many different papers.''} Others appreciated its ability to highlight structure, with P9 calling it \textit{``an interesting way of breaking down syntax''} and P14 noting that \textit{``the colors are so notably helpful to conceptualize the different components.''} For some, the tool’s comparative capabilities suggested clear applications: P7 pointed to its value for learning writing formats in legal research, while P12 considered the color-coding to be potentially \textit{``efficient and effective for understanding who said what.''}

\paragraph{Accessibility and usability considerations.}
Participants also stressed that the tool would need to be more accessible and usable in legal practice (P2, P7, P13, P18, P22). They called for clearer color differentiation in highlights (P2), support for right-to-left languages and color-blind accessibility (P7), and multilingual NLP to handle cross-border legal work (P13). Others emphasized the need to simplify the interface by hiding advanced features until needed (P18) and reduce information overload by showing fewer results at once (P22): \textit{``This feels a little bit like Google only giving 10 results; less would be more''} (P22).

\paragraph{Document types and applicability.}
Participants saw AbstractExplorer as especially useful for navigating large, disorganized legal documents like contracts and clause libraries, and for comparing legal language to spot patterns or anomalies. Others highlighted its potential in case law analysis and academic legal writing, where structure is more consistent. However, some were skeptical of its effectiveness with less formal or stylistically variable texts like memos or legal reviews. See ~\cref{app:ae_uses} for more details.


\paragraph{New directions inspired by AbstractExplorer.}
Many emphasized improved search and information retrieval, such as surfacing the precise sentence relevant to a legal issue (P1), cross-referencing terms across documents (P17), or flagging anomalies in clauses compared to common norms (P18). P9 suggested  \textit{``adding an inference layer to find contracts impacted by external events such as sanctions and insolvency,''} and P12 imagined a system that \textit{``made it clear to see who said what.''} Others suggested information tagging tools to automatically label contracts (names, dates, courts, risks) (P5), index contracts by section (scope, pricing, IP, confidentiality) (P3), highlight logical structures in clauses (P6), organize bookmarks into subcategories (P7), or label student writing (facts, law, application) for grading and feedback in law teaching (P22). Integration with existing firm infrastructure was also seen as critical, whether through export-to-table features for collaboration (P5) or embedding within document management systems like iManage or Relativity that stored client data (P5, P9).

\subsection{Writing Tool: CorpusStudio}
Many participants responded positively to the core idea of orienting writers to how others draft and structure similar texts (P6, P7, P11, P12, P14, P15, P17, P19, P20, P22). Some participants described the concept as “really interesting” and “useful,” or otherwise appealing for learning stylistic norms and drafting toward specific legal audiences (P6, P7, P17). P5 saw the tool, especially bookmarking and exporting, as useful for “junior training” and collaborative drafting handoffs. 

\paragraph{Comparisons with Existing Tools and Workflows}
Participants situated CorpusStudio relative to current practices and tools. P8 described in-Word assistants (e.g., DraftWise, Spellbook, Luminance) that surface firm and jurisdictional precedents alongside the active document; they framed CorpusStudio as conceptually similar but interface-distinct. P19 referenced Lexis and Westlaw as repositories of examples and styles across courts and firms, indicating an opportunity for CorpusStudio to complement research databases by showing how many writers phrase or structure similar material rather than only returning related sentences.

\paragraph{Usability, Learning Curves, and Cognitive Load}
The need for training and desire for low friction tools surfaced as a limitation. P3 reported that the workflow felt like “a lot of steps” and “not the act of writing,” noting the cognitive shift required to draft within the tool. P6 called the interface “a bit unintuitive” without a walkthrough. P21 said the writing experience in CorpusStudio could be “a little bit cumbersome,” particularly in commercial contracting contexts where writers typically start from templates and are expected to borrow and tweak language---because CorpusStudio prohibits copy-pasting from prior documents, in order to discourage plagiarism in the academic contexts it was originally designed for. 
Overall, they found it not directly useful for legal writing, but acceptable if positioned strictly as reference. Similarly, P11 noted that legal writing in firms often involves reusing prior work and adapting it with minimal changes, so a tool like CorpusStudio may not fit naturally into that workflow. However, they saw potential if it could help associates tailor their writing to match the stylistic preferences of different partners, which is a common but challenging part of firm practice. In direct contradiction with P21 and P11's emphasis on borrowing, P7 asked for built-in checks to detect when a draft is getting “too close” to sources.

\paragraph{Requested Controls and Feature Suggestions}
Participants asked for finer control over how content is organized and inspected. Requests included categories and sorting for saved notes (P7), filters oriented to audience and partner styles (P11), and support above the sentence level, e.g., ensuring logical transitions between sections (P20). 

Many also pointed to practicalities like the ability to upload and curate one’s own corpora (P14) and adaptive behavior that learns from re-ranking preferences (P19). P22 emphasized provenance as they initially would not use the tool (if pre-populated) for academic work without corpora transparency. P16 envisioned a self-populating and LLM-assisted corpus rather than only a fixed, pre-populated one. P6 discussed confidentiality, noting that unlike public research articles, most contract materials are private, suggesting increased value if firms could import internal databases.

The highlighting support drew mixed reactions. For example, P12 “love[d] the color coding” and found it “incredibly useful,” but P8 advocated for a more conventional redlining palette (red, blue, green) familiar to legal practitioners and widely used in law firm tooling.
 
\paragraph{Use Cases for CorpusStudio: Corpora to Query into While Writing}
Participants saw CorpusStudio as most useful for querying contracts, while emphasizing the need to preserve context within documents. They also highlighted its application to court oriented drafts, organization specific documents, and educational materials like student memos and exams. Additionally, some expressed interest in corpora for lay audience writing, including grade-level texts with guidance to serve as exemplars for clear communication. See ~\cref{app:corpusstudio_uses} for details. 

\paragraph{Expert Ideas for New Writing Support Tools}
Similar to AbstractExplorer, P17 valued CorpusStudio for enabling comparison and surfacing similar wordings during review. They emphasized the need for audience and venue tailoring with corpus control and hover over source inspection, ideally integrated with licensed research platforms like Westlaw or LexisNexis (P19). 

Participants also envisioned new writing support tools further afield from the prototypes that could support legal writing, such as section-level indexing and question-answering over long documents, to enable interrogating specific sections directly (P3, P6). They also requested to see multiple examples “in one steady view” to support section level navigation and synthesis (P14). Participants sought word processor integration with precedent retrieval and clause searching, or in-house drives to surface relevant precedents and clauses for side-by-side review (P8). One participant imagined scaffolds with click through provenance showing the source where language originated (P5). P9 envisioned contract mapping that instantly expands defined terms in place to avoid scrolling. P2 and P18 noted that a drafting assistant or mentor-style tool which mirrored the collaborative learning nature of junior and senior review would be useful. P16 requested predictive sentence completion with optional, source backed, rationale on why the option was suggested. 

For drafting and review, like CorpusStudio, they sought contextual retrieval of similar agreements and clauses to compare phrasing and understand differences (P8), as well as advanced search that stacks local context windows with bookmarking and term co-occurrence insights (P12), and robust comparison/“diff” workflows: for example, perhaps one could drag an email containing requested clause changes, automatically compare against standard terms, highlight and bullet key differences, and emit a version controlled Word document aligned to a chosen risk profile (P21). Participants using document automation and template management systems wanted features to support built in contextual research (P2). 

Lastly, participants cautioned against over reliance on tools. Even when framed as decision support, as P22 noted, tools can drift into decision making. They suggested design guardrails to help preserve human oversight in tasks like exam grading or literature review, where there is a risk of citing without reading (P22). This concern intersects with broader calls for auditable systems, clearer provenance, and interfaces that encourage informed judgment rather than passive acceptance (P7, P18).

\subsection{Additional desires for future systems}
Several participants articulated a need for more agentic AI that automates routine workflows (P11, P16, P22). As P11 put it, this would mean building for the ``AI-powered professional,'' where the system handles repetitive tasks like uploading, prompting, and scheduling review, while leaving substantive judgment to the lawyer. Others emphasized that such systems should fine tune or refine legal opinions but never replace professional expertise (P16).

\subsection{Strengths and Gaps in Existing Tools}
\subsubsection{Reading Support}
Participants described recurring pros and cons in the reading tools they use. Several praised aggregation and “getting up to speed” affordances: for example, P2 relies on Feedly to centralize monitoring and liked its AI-driven feeds for surfacing sources they were not previously aware of. P3 highlighted core affordances such as Ctrl-F and using AI summaries as a “Table of Contents” style starting point before deeper reading. P20 similarly pointed to low investment utilities (e.g., spell check) that deliver quick, dependable gains. 

Others emphasized comparison, difference finding, and document logic. P4 framed traditional redlining as “okay” only when differences are small; with many changes it “breaks down” and lacks a richer “third dimension.” They argued for git-style versioning and collaboration to focus on “what’s different” and deal points, essentially reflecting frustration with path-dependent, Word-centric workflows. 
P15 praised AI’s speed at summarizing across large aggregates and its “compare and contrast” capabilities, but distrusted outputs when documents interleave multiple terms and dates, or list many termination routes; they described these as cases where simple prompts can return the wrong answer and there was a need for more sophisticated prompts. 


Accuracy, model transparency, and prompting burden were persistent concerns. 
For example, P7 was uneasy about models being opaque, yet values AI-support tools as proofreaders to simplify and transcribe text and flag inaccessible language. P9 uses AI to summarize and to explain unfamiliar terms “in the context of [their] market,” but stressed that extraction and summarization requires detailed instructions; they described that AI-support tools work well for high volume, repeat matters, and less so when every document is novel. They described breaking tasks into explicit, step-by-step prompts that point to specific sections, and once set up, scaling review across many documents while still performing human checks. P15 echoed this sentiment, saying there were use cases where simple prompts can return the wrong answer, so they thought there was a need for more sophisticated prompts.

Control over sources and corpus transparency also mattered. P17 said hallucination risk makes a full skim “unavoidable,” preferring tools that still let them read end-to-end. They would trust outputs more if they could supply and control the underlying database (like CorpusStudio allows). 

Availability, integration, and interface fit were also discussed as limitations of current tools. P8 reported not using reading support tools and criticized how many offerings are enterprise or Microsoft Word plugins without Google Docs support, wishing for a Chrome or Safari extension with the same functionality. P19 underscored the friction of multipane interfaces and asked for a single condensed interface with filters that supports one-pass confidence when goals permit (tagging vs.~first-and-only review). They noted screen real estate and device constraints (phone versus desktop) and the importance of interfaces that “auto fit” to the hardware and task.

\subsubsection{Writing Support}
Traditional utilities like spell check and Ctrl-F/Find-and-Replace were seen as both useful and intrusive, old and out-dated or simple and powerful. Spell-check can continuously bother a participant about correctly spelled words or minor suggestions (P3), while another participant valued it precisely because it delivers gains with minimal investment (P20). P12 thought Ctrl-F/Find-and-Replace was insufficient for context aware filtering and intelligent result selection, while P3 thought of it as a valuable core affordance. 

AI-based tools were appreciated for speed, jump-starting drafts, and relieving the blank page problem. Participants used systems like Harvey for initial summaries and first drafts, helping teams get up to speed (P5), and relied on in-house models for targeted tasks such as rephrasing (P9). Others noted that AI can quickly provide structure, templates, and connective prose between points (P15) or streamline routine copyediting and readability checks via Hemingway or Grammarly (P7). One emphasized that general-purpose chat tools like ChatGPT and Perplexity are more “sophisticated:” they stay out of the way until invoked, and can search, write, and format tables on demand (P3). Grammar and style cleanup that improves clarity and polish (sometimes at the cost of personal voice) was seen as a benefit, with personalization improving as the system ingests more of the participants writing (P19).

At the same time, participants described limitations around accuracy, domain specificity, transparency, and integration into legal workflows. Some found suggestions unhelpful because they miss document context or preferred structure (P9), or because models are not tuned to legal language and style (P6). Others flagged inconsistent answers that push users to retry prompts repeatedly (P12) and knowledge cutoffs that limit up-to-date responses (P17). 
Participants wanted to understand why an answer was produced and felt current “explanations” remain too high level (P7, P18). Interface behavior mattered: similar to the annoyance with spell-check being intrusive, Copilot was perceived as invasive and “butting in;” there was a preference for optional suggestions exemplified rather than imposed (P3). Even when AI accelerates work, participants stressed the need to double check outputs, with one participant noting that “speed doesn’t mean quality", and continued by describing organizational frictions: tool complexity and ongoing training, junior and senior colleagues tool adoption gaps, privacy and licensing considerations, budget and pilot training constraints, and even economic disincentives under billable hour models (P5). Echoing budget consideration, another participant mentioned that subscription and enterprise licensing models also constrained access (P6). 

\subsection{Where AI Stops: Tasks Kept for Humans}
Participants drew clear boundaries around tasks that should remain under human control, particularly final authorship and review. Many would not delegate final legal writing to AI, even if it could produce strong drafts. They emphasized the need for precise, accessible, and “bulletproof” language, as well as personal responsibility (P1, P12, P15, P19). Others stressed that a human must review outputs before submission, citing ethics and risk (P4, P6, P17). One resisted relying on AI to read judicial opinions, noting that summaries can miss nuance and obscure argumentative opportunities, while also expressing concern over losing their own voice in writing (P14). 

Strategic, relational, and judgment-heavy tasks were also reserved for humans. Participants rejected automating client management and trust-building, describing justice as a relationship rather than a transaction, and noting risks like deepfakes (P5). Sharing findings with clients or counterparties was framed as inherently strategic, even when facts were correct (P9). P16 argued that goal-setting, tradeoffs, and client-specific priorities are core legal functions unsuited to automation.

Operational boundaries also mattered. Participants resisted letting AI finalize or send outputs without human oversight. For example, sending a contract with unvetted edits was seen as unprofessional (P2). Some avoided tools like Copilot to protect critical records (P3). Concerns over hallucinations, unpredictable prompts, and the impossibility of pre-specifying checks reinforced the need for human review (P17). One limited AI use to edits, not full drafting, to preserve authorship and voice (P19).

Not all participants drew strict limits. A few said there was nothing they categorically wouldn’t let AI do, so long as human oversight and context-setting were present. One saw AI use as an ethical obligation to clients, another cited EU rules requiring human control, while endorsing broad use with review, a third emphasized iterative improvement via better prompting and audience awareness (P11, P18, P21). Outside legal practice, one insisted on designing lectures personally to preserve sincerity and meet student expectations (P22).

\subsection{Factors Considered Before Tool Adoption}
Participants cited clear prerequisites for adopting tools. Chief among them were data protection and compliance. Several emphasized confidentiality, data storage within compliant jurisdictions, and prohibitions on model training over their data (P6, P7, P8, P12, P15, P19). Regulated environments required institutional control over data (P7). Others wanted high-level understanding of the tool’s design, sources, and model configuration (e.g., temperature) (P14, P17, P18). Legal exposure was also a concern, including liabilities from cached outputs (P15). 

Trust depended on accuracy, reliability, and verifiability. Participants repeatedly cited these as core requirements, along with effectiveness (P3, P9, P12, P13, P16). One wanted systems that linked answers to exact source references (P9). Some tested tools by asking known questions or comparing outputs to prior work (P11, P15). One preferred retrieval-augmented generation over generalist models, favoring best-in-class tools for specialized tasks (P11).

Ease of use and workflow impact were equally important. Participants favored intuitive, fast tools with little or no learning curve, minimal clicks, and measurable efficiency gains, especially under time pressure (P3, P7, P9, P17, P20, P21). One required “extremely user-friendly” tools that reduced repeated legwork (P21). Another accepted slower tools if they improved quality, but emphasized efficiency as central to practice (P19).

Cost, trials, and social context also influenced decisions. Participants sought free trials, low upfront costs, and clear return on investment (P8, P11). P21 stressed the need for a business case given high enterprise prices. Social proof, hearing that peers used and liked a tool, increased adoption (P8). Team culture, leadership support, and training promoted uptake, while their absence hindered it (P5). One followed structured pathways from awareness to piloting (P7) and another relied on trial-and-error (P22). One participant rejected vendor locked systems, preferring open, modular tools to avoid long-term rigidity (P4).

\subsection{Concerns Around Tool Use in Legal Work}
Participants consistently flagged confidentiality and data privacy as top risks. One emphasized protecting client, state, and company materials from breaches, warning that a hack would be “detrimental” (P1). Another noted that leakage could expose not only documents but also negotiation patterns and internal thinking (P8). P9 required “top-end enterprise-level security,” including bans on training external models with their data. One described formal IT vetting as part of organizational safeguards (P21).

Accuracy concerns, especially overlooked risks or hallucinations, were seen as intolerable (P2, P16). One noted that hallucinations add to review time, and current tools do not reliably flag errors; they did not trust AI to self-check and insisted on human verification (P17).

Participants also raised concerns about provenance and integrity. One regularly asked the support tool to provide sources and noted inconsistencies across platforms, undermining trust (P3). P14 echoed this, emphasizing the need to understand what data is used and question whether systems are truly closed.

Integration modalities and user control also mattered. One participant viewed embedded, always-on assistants (e.g., copilots in existing work surfaces) as “more risky” than deliberately accessing tools, warning the potential for undetected detours in the workflow that could be hard to detect (P3). 

Not all participants shared these concerns. One participant, self-described as “a fan of AI,” reported no added worries (P18).

\section{Discussion} 
Using AI-resilient text-preserving probes such as GP-TSM for source-anchored skimming, AbstractExplorer for cross-document comparison, and CorpusStudio for corpus-guided drafting let participants show, in context, both what they wanted and where they drew hard lines. GP-TSM clarified the central tension: participants welcomed faster orientation so long as the original wording stays visible and verifiable; many preferred it over black box summaries precisely because they could verify what was de-emphasized, reducing fear of omissions or hallucinations. At the same time, several insisted it is unsuitable for first reads or for high-stakes texts where lexical shifts and every word matter. 

AbstractExplorer exposed a complementary hope: tools that help experts see structure and patterns across many documents at once, useful for scanning long, poorly organized repositories, comparing clause language, and learning format conventions, while also surfacing practical constraints (accessibility, overload) that would need design attention. 

Lastly, CorpusStudio showed that precedent-based legal writing brings both benefits and challenges. Participants liked orienting to how others structure and phrase similar texts and asked for provenance, integration with tools they already use, and controls keyed to audience/style. But they also flagged workflow costs (extra steps), template heavy habits (copy/tweak), and the need to preserve intra-document context when retrieving smaller units, highlighting important considerations for AI tool design. As such, we have the following design implications for future novel legal interface development:

\paragraph{Integrate into Existing Workflows with Granular Control.}
Participants emphasized the importance of embedding new features directly within familiar tools like Word or Adobe, reducing training overhead and supporting seamless adoption. Lightweight yet powerful enhancements, such as skimming aids or context-aware overlays, should offer fine grained user controls to align with existing practices without overwhelming them.

\paragraph{Design for Verifiability and Trust through Transparent Provenance}
Legal professionals stressed the need for outputs to be traceable back to specific source texts. They preferred retrieval grounded responses over opaque model outputs and desired mechanisms to test outputs against known documents, supporting verification, audit-ability, and trust.
\paragraph{Adapt Communication for Risk and Audience Sensitivity}
Participants wanted systems that reflect varying levels of risk and audience needs, offering internal views for fast comprehension and externally appropriate summaries for sharing. Features that highlight changes from past decisions or redlines were also seen as critical for maintaining accountability and transparency.
\paragraph{Balance Speed with Confidentiality}
While speed and efficiency were valued, participants were clear that these must not come at the cost of lost confidentiality. Interfaces should remain low friction and time saving, especially in high-stakes environments, but still uphold strict privacy standards.

\section{Conclusion}
Law is one domain where small words, punctuation, and cross-references carry consequences; several participants cautioned that for high-stakes texts they must read by the letter, indeed, a setting where every word matters. Comparable stakes appear in other domains such as medicine, aviation, and finance, where minor lexical shifts change actions, safety, or liability. The insights and patterns that we surface could offer transferable value for these fields. Exploring the application of these insights in other high-stakes fields could lead to broader improvements in AI-assisted support tools.





\bibliographystyle{ACM-Reference-Format}
\bibliography{main}

\appendix
\section{Pre-study Survey}
\label{app:pre-study}
\subsubsection{Demographics} 
\begin{enumerate}
    \item What is your participant ID?	\newline
    (Provided to participant by interviewee beforehand)
    \item What is your age? \newline
        \begin{itemize*}[label=\textopenbullet]
            \item 18-24
            \item 25-34
            \item 35-44
            \item 45-54
            \item 54+
            \item Prefer not to disclose
        \end{itemize*}
    \item What is your gender?	\newline
        \begin{itemize*}[label=\textopenbullet]
            \item Woman
            \item Man
            \item Non-binary
            \item Prefer not to disclose
            \item Other
        \end{itemize*}
\end{enumerate}

\subsubsection{Experience} 
\begin{enumerate}
    \item What is your current professional role or title?
    \item How many years of professional experience do you have in your current field? Use a numeric answer.
    \item How many years of experience do you have with \textbf{reading} legal documents? Use a numeric answer.
    \item How many years of experience do you have with \textbf{writing} legal documents? Use a numeric answer.
\end{enumerate}

\subsubsection{Frequency of writing and reading legal documents} 
\begin{enumerate}
    \item How often do you \textbf{write or revise} legal documents in your work?
        \begin{itemize*}[label=\textopenbullet]
            \item Never
            \item Rarely
            \item A few times a month
            \item A few times a week
            \item A few times a day
            \item Other
        \end{itemize*}
    \item How often do you \textbf{read} legal documents in your work?
        \begin{itemize*}[label=\textopenbullet]
            \item Never
            \item Rarely
            \item A few times a month
            \item A few times a week
            \item A few times a day
            \item Other
        \end{itemize*}
\end{enumerate}

\subsubsection{Tool Usage} 
\begin{enumerate}
    \item Which of the following types of tools have you used in your work? \textit{(Select all that apply)}
        \begin{itemize*}[label=\textopenbullet]
            \item Microsoft Word/Google Docs
            \item AI assistants (e.g. ChatGPT, Copilot, Claude)
            \item Legal/clinical research tools (e.g. Westlaw, LexisNexis)
            \item Writing enhancement tool (e.g. Grammarly, Wordtune)
            \item None
            \item Other
        \end{itemize*}
    \item How often do you use support tools for \textbf{writing}?
        \begin{itemize*}[label=\textopenbullet]
            \item Never
            \item I've used tem a few times, but not regularly
            \item A few times a month
            \item A few times a week
            \item Once a day or more
        \end{itemize*}
    \item How often do you use support tools for \textbf{reading}?
        \begin{itemize*}[label=\textopenbullet]
            \item Never
            \item I've used tem a few times, but not regularly
            \item A few times a month
            \item A few times a week
            \item Once a day or more
        \end{itemize*}
\end{enumerate}

\subsubsection{Current Practices} 
\begin{enumerate}
    \item When \textbf{reviewing} long documents, how often do you rely on skimming rather than reading in detail?
        \begin{itemize*}[label=\textopenbullet]
            \item Never
            \item Rarely
            \item Occasionally
            \item Often
            \item Very often
        \end{itemize*}
\end{enumerate}

\subsubsection{Acknowledgment of Participation} 
\begin{enumerate}
     \item Would you like to be acknowledged by name in any future publications that result from this research (e.g., in the Acknowledgments section)?
        \begin{itemize*}[label=\textopenbullet]
            \item Yes, include my name.
            \item No, do not include my name.
        \end{itemize*} 
\end{enumerate}

\section{Background and Current Practices}
\label{app:bacp-questions}
\subsubsection{Reading}
\begin{enumerate}
    \item What types of documents do you typically read (e.g., memos, briefs, motions, rulings, contracts, intake summaries)?
    \item What do you consider the most challenging part of reading legal documents in your role?
    \item How do you typically approach reading a legal document? 
\end{enumerate}

\subsubsection{Writing}
\begin{enumerate}
    \item What types of documents do you typically write (e.g., memos, briefs, motions, rulings, contracts, intake summaries)?
    \item What do you consider the most challenging part of writing legal documents in your role?
    \item How do you typically approach reading a legal document (where do you start? What references or templates do you use if any)?
\end{enumerate}

\section{Support Tool Questions}
\label{app: tool-qs}
\subsubsection{GP-TSM}
\begin{enumerate}
    \item What are your initial thoughts on this tool?
    \item What would you want to skim or read at scale?
    \item Given that this tool exists, what type of tools would be useful within your context?
\end{enumerate}

\subsubsection{Abstract Explorer}
\begin{enumerate}
    \item What are your initial thoughts on this tool?
    \item What corpora would you want to skim or read at scale?
    \item Given that this tool exists, what type of tools would be useful for exploring document collections in your work?
\end{enumerate}

\subsubsection{Corpus Studio}
\begin{enumerate}
    \item What are your initial thoughts on this tool?
    \item What corpora would you use with a tool like this (What type of documents would you want to query into when writing)?
    \item Given that this tool exists, what type of tools would be useful for looking into particular parts or sections of longer documents? [ability to index into longer documents]
\end{enumerate}

\section{Open-Ended Interview Questions}
\label{app:open-questions}
\begin{enumerate}
    \item What are some of the pros and cons of tools that you currently use for writing (non AI based and AI based)?
    \item What are some of the pros and cons of tools that you currently use for reading (non AI based and AI based)
    \item What’s the most challenging task for you that you, even with its complexity, wouldn't want AI to do for you?
    \item What factors do you consider before adopting a tool into your workflow? 
    \item Are there risks or sensitivities you'd be concerned about if using reading and writing support tools in your work (e.g., confidentiality, bias, citation accuracy, hallucinated content)?
\end{enumerate}

\section{Types of Documents Read and Written by Legal Experts}
\label{app:docs-read-written}
\subsubsection{Reading}
Participants described engaging with a  wide spectrum of legal texts in their day-to-day work. Some emphasized formal legal sources such as case law (P1, P2, P6, P8, P9, P22), judgments (P1, P7), statutes and amendments (P1, P6, P15, P21), and regulatory filings (P12, P13, P15, P20, P22). Others focused on litigation documents, including motions, affidavits, trial transcripts, and pleadings (P5, P10, P19, P20, P22). Many participants highlighted contracts, agreements, NDAs, and corporate documents as central to their practice (P3, P4, P5, P8, P9, P11, P15, P16, P17, P18, P21). Everyday communication such as emails, memos, internal policies, and reports also constituted a significant portion of their reading load (P3, P6, P14, P17, P19, P20, P21). Others engaged with law review articles and scholarly research (P2, P7, P13, P14, P15, P22), training materials (P2), newsletters (P14), and texts such as marketing material or website disclaimers (P21). This list showcases the wide variety of reading materials used across different legal roles and domains.

\subsubsection{Writing}
Within their workflows, participants drafted legal documents spanning litigation, transactional, advisory, organizational, and scholarly outputs. Litigation work included head of arguments (P1), affidavits (P1), motions (P5, P10), complaints (P10), legal briefs (P3, P5, P11, P17, P20), user-facing court documents (P7), legal opinions (P6, P17), claims (P17), and legal reports (P17). Transactional documents included contracts (P2, P3, P4, P5, P6, P9, P10, P11, P15, P16, P17, P18), NDAs (P4), agreements (P11, P15, P16), transactional documents (P4), negotiation-related docs (P11), risk summaries (P4), and contract templates (P16). Advisory and policy outputs included summaries of regulations (P12), guidelines (P2, P7), policy summaries/white papers (P13, P18), and regulations (P20). Communication and organizational documents ranged from emails (P3, P10, P14, P17, P20) and memos (P14, P19, P20, P21) to PowerPoints (P2, P21). Participants also authored case studies (P3), forms (P7), rule books (P9), books (P13), academic papers or law articles (P13, P14, P22), and journalistic pieces or op-eds (P14), alongside routine revisions to existing documents (P5, P8, P15, P16, P19) and revisions (P6).

\section{Use cases for GP-TSM}
\label{app:gptsm-uses}
Participants converged on several use cases where a skimming support or AI-resilient summarization tool like GP-TSM could be helpful, falling broadly into three categories: overly long or formulaic documents, formal or academic reading materials, and practice-facing communications. Overwritten or formulaic documents included legal contracts, consumer complaints, and legal deeds, often described as repetitive and unnecessarily lengthy (P3, P5, P8, P16, P17, P18, P22). For example, one participant remarked that “an 18-page complaint… could have been a page and a half” (P3). Others noted long records such as trial transcripts, technical product documentation, and regulatory guidelines (P5, P15, P18) as additional candidates for summarization. Formal and academic materials such as research articles (P6, P7, P15), judicial or legal opinions (P6, P14, P19), case studies (P7, P19), regulatory publications (P9), executive summaries, and government filings (P12) were also seen as well-suited for GP-TSM, particularly to surface main takeaways more efficiently. Practice-facing documents and communications included opposing counsel’s briefs (P10, P12), client communications and new case-relevant content (P9), long emails (P19), expository legal memos and court briefs (P16, P20), and law school materials like casebooks (P22). One participant noted they would apply GP-TSM “across all document types” to aid in reading (P16), while another mentioned they would use it in contexts “where losing nuance is not critical,” such as summaries or synopses (P13).

Some participants were explicit about documents they would not use GP-TSM for. Participants warned against using GP-TSM for texts where every word matters, such as contracts, statutes, and regulatory documents (P2, P6, P11, P13, P14, P15, P20). For example, P15 explained, “the phrasing can… turn an entire sentence… skipping over those could be kind of fatal to an argument,” and P13 emphasized that legal texts must be read “by the letter.” P14 added that when documents are cited, they felt “ethically and… substantively compelled to read the whole thing.” P20 raised similar concerns about statutes and how GP-TSM would handle legal citations in briefs. Others noted that GP-TSM might be inappropriate depending on the situation, particularly for first time readings, litigation scenarios, or when dealing with adversaries (e.g., “when the counterpart is naughty,” P18), where missing nuance could be consequential (P10, P18). Some also felt it was simply unnecessary for simple or easily skimmable texts (P8). One participants flagged GP-TSM as possibly useful for students or casual reading, but not for professional legal work (P11).

The findings show that while some participants supported the use of GP-TSM for certain documents, others expressed reservations about those same types. These diverging opinions were often rooted in differing perceptions of risk, context, and the tradeoff between efficiency and precision. Some viewed GP-TSM as an asset for enhancing speed and usability, while others saw it as a risk due to the importance of preserving nuance and legal fidelity.

\section{Use cases for AbstractExplorer}
\label{app:ae_uses}
Participants identified several distinct use cases where AbstractExplorer showed strong potential. Many saw particular value in using the tool to navigate large or poorly organized legal documents, such as contracts or clause libraries (P3, P6, P9, P12, P13, P15, P17, P18). For example, P3 described how the tool could serve as an “objective summary, essentially indexing a 40–60 page contract,” while P9 highlighted its ability to “give a high-level overview and categorize what’s in your document management system.” Others emphasized its usefulness for identifying patterns and comparing language across documents, such as P17’s suggestion to compare arbitration clauses to spot “common vs. uncommon wording,” or P18’s idea of checking whether a clause aligns with standard practices or stands out as a “Black Swan.” A second theme focused on the tool’s potential in case law and litigation workflows (P1, P9, P15, P19), including pinpointing key sentences in rulings (P1, P15) and analyzing stylistic tendencies in briefs and arguments (P9). A third group saw promise in academic and scholarly legal writing, such as peer-reviewed law journals, which share structural similarities with scientific abstracts (P7, P12, P14, P20, P22). At the same time, participants expressed skepticism about applying the tool to less structured legal texts, like memos or legal reviews, which often feature rhetorical or inconsistent writing. As P22 noted, “Legal review is florid and bombastic, unlike scientific writing, so not sure that kind of writing would play so nice with this.”

\section{Use Cases for CorpusStudio}
\label{app:corpusstudio_uses}
Participants identified a variety of corpora they would query into using CorpusStudio, with its utility depending largely on the type of document. Many highlighted contracts as the primary corpus due to their length, repetitive structure, and template-like conventions (P2, P3, P5, P6, P8, P9, P13, P15, P16, P20, P22). While some favored highly standardized contracts, others preferred applying the tool to less uniform contracts to better navigate heterogeneity. A common concern was the importance of preserving intra-document context when retrieving smaller sections from contracts. Beyond contracts, participants proposed corpora for narrative, court-oriented drafting, including statements of claim, complaints, pleadings, briefs, and legal opinions (P5, P6, P9, P15, P16, P19, P22), especially where document structure is conventional or expected by authorities. Several participants emphasized querying their own organization’s documents, such as firm playbooks and negotiation parameters, to maintain consistency and institutional style across similar matters (P2, P3, P9, P17). Other corpora included standardized or reference-heavy materials, like organizational policies, public documents, and court transcripts (P5, P15, P16). For educational contexts, suggested corpora included student legal memos and exams to support teaching and comparative review (P14, P22), while academic papers were recognized as a distinct writing corpus (P7, P19). To aid writing for lay audiences, participants also expressed interest in corpora written at grade school reading levels, complemented by guidance from design groups to serve as exemplars for clear communication (P7).

\end{document}